\documentclass{article}
\usepackage{spconf,amsmath,graphicx}
\usepackage{amssymb}
\usepackage{multirow}
\usepackage{subfig}
\usepackage{wrapfig}

\title{ETP: Learning Transferable ECG Representations via ECG-Text Pre-training}
\name{Che Liu$^{1, \star}$, Zhongwei Wan${^{2, \star}}$\thanks{$\star$ Equal Contribution.}, Sibo Cheng$^{1}$, Mi Zhang$^{2}$, Rossella Arcucci$^{1}$}
\address{
    $^1$ Imperial College London, UK \\
   $^2$ The Ohio State University, USA\\
    \small{
        \{che.liu21, sibo.cheng, r.arcucci\}@imperial.ac.uk,
       \{wan.512,  mizhang.1\}@osu.edu
    }
}
\begin{document}
%
\maketitle
\begin{abstract}
In the domain of cardiovascular healthcare, the Electrocardiogram (ECG) serves as a critical, non-invasive diagnostic tool. Although recent strides in self-supervised learning (SSL) have been promising for ECG representation learning, these techniques often require annotated samples and struggle with classes not present in the fine-tuning stages. 
To address these limitations, we introduce ECG-Text Pre-training (ETP), an innovative framework designed to learn cross-modal representations that link ECG signals with textual reports. For the first time, this framework leverages the zero-shot classification task in the ECG domain. ETP employs an ECG encoder along with a pre-trained language model to align ECG signals with their corresponding textual reports. The proposed framework excels in both linear evaluation and zero-shot classification tasks, as demonstrated on the PTB-XL and CPSC2018 datasets, showcasing its ability for robust and generalizable cross-modal ECG feature learning.
\end{abstract}
\section{Introduction}
The Electrocardiogram (ECG) is a crucial clinical diagnostic tool for various cardiac conditions. While deep learning has shown promise in ECG classification, its effectiveness often depends on the availability of high-quality labels and expert review, making the process labor-intensive and costly.

\begin{figure*}[htp!]
\centering
 \subfloat[Zero-shot classification pipeline.\label{2a}]{\includegraphics[width=0.45\linewidth]{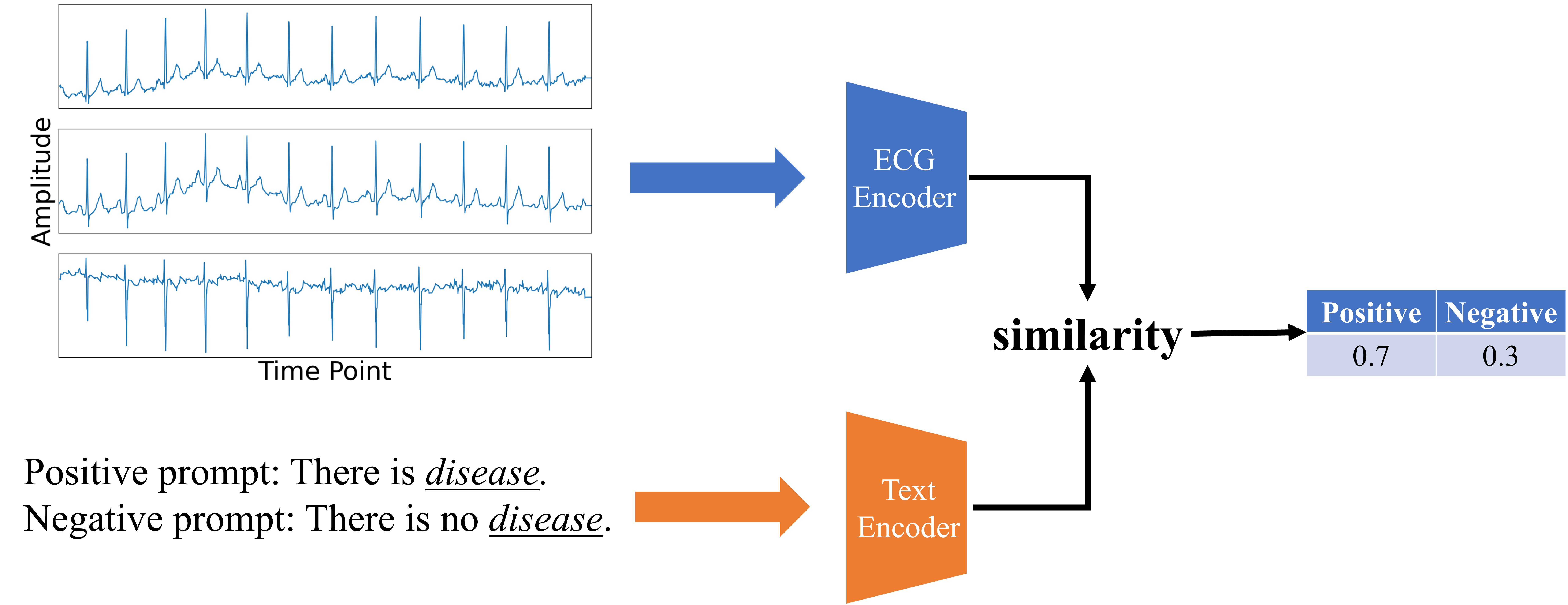}}
 \hfill
\subfloat[Fine-tune classification pipeline\label{2b}]{\includegraphics[width=0.45\linewidth]{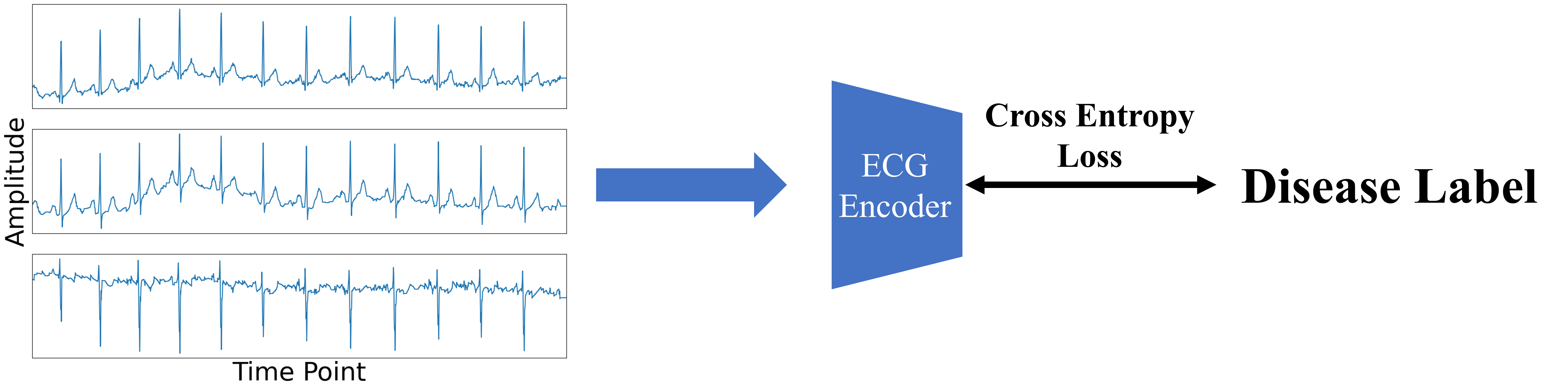}}
    \hfill
\caption{The pipeline of zero-shot classification and fine-tune classification.}
\label{fig2: downstream}
\end{figure*}

In the quest to circumvent the pitfalls of extensive annotation, self-supervised learning (SSL) has emerged as a promising avenue, excelling particularly with datasets harboring limited annotations~\cite{jaiswal2020survey,chou2020knowledge,wan2023med}. SSL paves the way for harnessing ECG representations beneficial for a spectrum of downstream tasks like abnormality detection and arrhythmia classification~\cite{liu2023spectral,mehari2022self}. However, a significant bottleneck remains: extant ECG SSL \cite{astcl,tstcc,chen2023learning,chenself} strategies still lean heavily on substantial annotated data for fine-tuning on downstream applications as shown in Fig \ref{2b}. Such a dependency becomes particularly limiting for rare cardiac conditions, steering research attention to the zero-shot classification. This paradigm aims to negate the need for annotated samples of unseen categories by leveraging cross-modal representation from ECG and disease-related textual prompt and utilize the ECG-text similarity to determine the predicted disease that do not need annotated data in downstream tasks as depicted in Fig \ref{2a}. 

The path to zero-shot learning for ECG isn't devoid of obstacles. Primarily, there exists a semantic disjunction between the continuous numerical nature of ECG and the discrete clinical terminologies in textual reports~\cite{li2023frozen,liang2022foundations,gtgm}. Further complications arise from domain adaptation issues and scalability concerns, with zero-shot models often requiring considerable computational resources~\cite{liu2023m}. While recent studies, such as those by~\cite{yamacc2022personalized} and~\cite{bhaskarpandit2022lets}, have made headway in ECG zero-shot classification, they remain tethered to supervised learning during pre-training, demanding extensive annotated ECG data.

Witnessing the potential of vision-language pre-training in broader contexts, as evidenced by works like CLIP~\cite{radford2021learning}, we introduce \textbf{ECG-Text Pre-training (ETP)}. This innovative approach seeks to leverage the 12-leads ECG and its corresponding textual reports within a cross-modal learning paradigm. ETP features a language model paired with an ECG encoder to yield text and ECG embeddings. Leveraging a priori clinical knowledge, the text is channeled through a sizeable frozen language model with 1D CNN serving as the ECG encoder's backbone. Both components possess linear projection heads, ensuring the harmonization of text and ECG dimensions. Following this, the concordance between ECG and text embeddings becomes the focal point to minimize contrastive learning loss and yield classification probabilities for diverse ECG categories.

The key contributions from our research are outlined as follows:
\begin{itemize}
\item We are the pioneers in delving into and unveiling the potential of ECG-Text Pre-training (ETP) specifically for ECG signals.
  
\item Our approach not only achieves state-of-the-art (SOTA) results in the fine-tuning phase but also becomes the first to demonstrate the viability of zero-shot tasks. Furthermore, compared to uni-modal SSL, our method exhibits enhanced robustness and transferability.

\item We have established the comprehensive benchmark for ETP, focusing on the confluence of ECG-Text pre-training and ECG signals.
  
\end{itemize}

\section{Methodology}

\subsection{ECG-Text Pre-training}

\begin{figure*}[htp!]
\centering
 \subfloat[ETP learns cross-modal ECG representations from paired ECG signals and clinical reports.\label{1a}]{\includegraphics[width=0.45\linewidth]{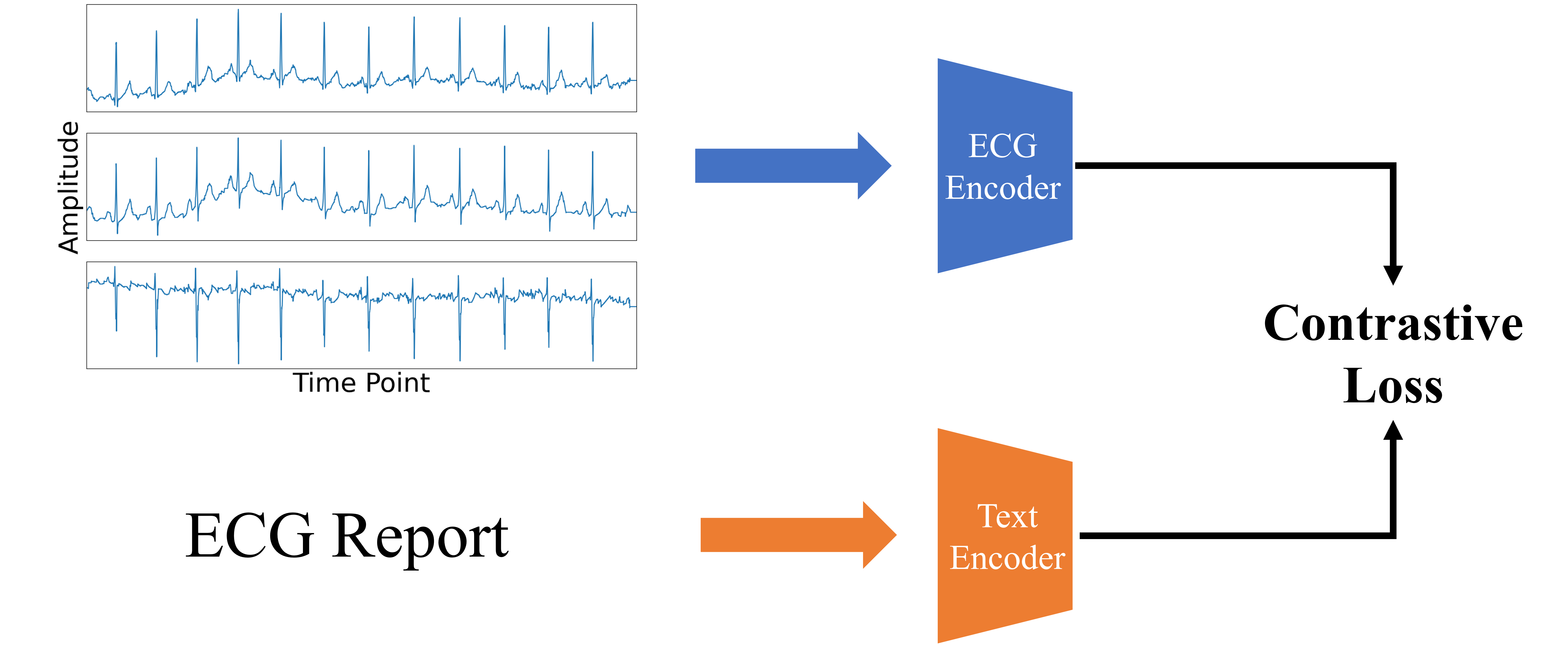}}
 \hfill
\subfloat[SSL learns ECG representations by maximizing the similarity between two unique augmented views.\label{1b}]{\includegraphics[width=0.45\linewidth]{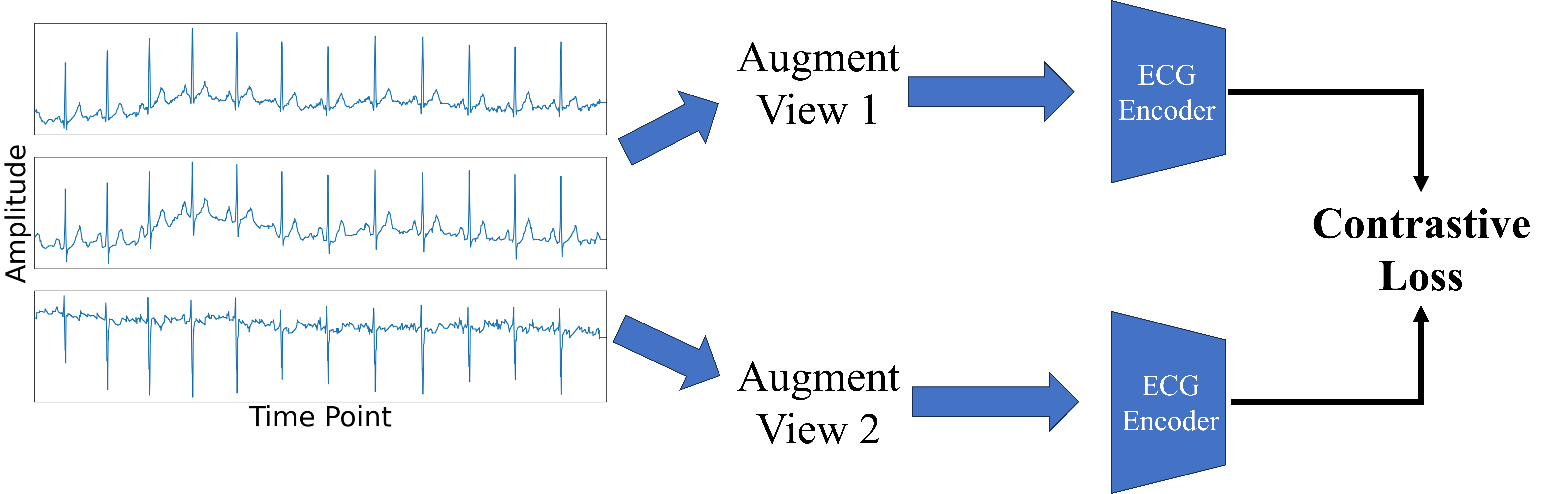}}
    \hfill
\caption{Comparison between ETP and SSL.}
\label{fig1: framework}
\end{figure*}

Incorporating both ECG signals and paired textual description, we employ the following modifications based on the CLIP framework:

Given the CLIP framework as a reference~\cite{clip}, we integrate a contrastive learning aiming to predict the associated pair $(e_{ecg, i}, t_{ecg, i})$ among the $N \times N$ probable ECG-text combinations, while strategically positioning the remaining $N^{2}-N$ negative combinations at a distance. 
In this context, two distinct encoders for ECG signals and text, denoted as \( \mathcal{F}_{ecg} \) and \( \mathcal{F}_{text} \) respectively, transform \( \mathbf{e}_{ecg, i} \) and \( \mathbf{t}_{ecg, i} \) into a latent embedding space, represented as \( [\mathbf{\hat{e}}]_{i=1}^{N} \). Subsequently, two separate non-linear projectors for ECG signals and text, denoted as \( \mathcal{P}_{ecg} \) and \( \mathcal{P}_{text} \) respectively, convert \( \mathbf{e}_{ecg, i} \) and \( \mathbf{t}_{ecg, i} \) into a consistent dimension, termed \( d \).
This process can be represented as:

\begin{align}
\hat{\mathbf{e}}_{ecg, i} &= \mathcal{P}_{ecg}(\mathcal{F}_{ecg}(\mathbf{e}_{ecg, i})), \\
\hat{\mathbf{t}}_{ecg, i} &= \mathcal{P}_{t}(\mathcal{F}_{text}(\mathbf{t}_{ecg, i})),
\end{align}

with both $\hat{\mathbf{e}}_{ecg, i}$ and $\hat{\mathbf{t}}_{ecg, i}$ belonging to the set $\mathbb{R}^{d}$.

From the training set, we extract ECG feature vectors denoted by $[\mathbf{\hat{e}}_{ecg, i}]_{i=1}^{N}$ and text feature vectors represented by $[\mathbf{\hat{t}}_{ecg, i}]_{i=1}^{N}$. Following this, we then calculate the cosine similarities as $r_{i,i}^{e2t} = \mathbf{\hat{e}}_{ecg, i}^{\top} \mathbf{\hat{t}}_{ecg, i}$ and $r_{i,i}^{t2e} = \mathbf{\hat{t}}_{ecg, i}^{\top} \mathbf{\hat{e}}_{ecg, i}$, which illustrate the ECG-text and text-ECG compatibilities respectively. The loss function, $\mathcal{L}_{\mathrm{CE}}$, is then expressed as:

\begin{align}
\mathcal{L}^{e2t}_{e} = -\log \frac{\mathrm{exp}(r_{i,i}^{e2t} / \sigma_{1})}{\sum_{j=1}^{K}{\mathrm{exp}(r_{i,j}^{e2t} / \sigma_{1})}}, \\
\mathcal{L}^{t2e}_{i} = -\log \frac{\mathrm{exp}(r_{i,i}^{t2e} / \sigma_{1})}{\sum_{j=1}^{K}{\mathrm{exp}(r_{i,j}^{t2e} / \sigma_{1})}}
\end{align}

\begin{equation}
\mathcal{L}_{\mathrm{total}}=\frac{1}{2 K} \sum_{i=1}^N\left(\mathcal{L}^{e2t}_{e}+\mathcal{L}^{t2e}_{t}\right), 
\end{equation}

Here, $\mathcal{L}^{e2t}_{e}$ and $\mathcal{L}^{t2e}_{t}$ are ECG-text and text-ECG cross-modal contrastive losses respectively. $\sigma_{1}$ denotes the temperature hyper-parameter, which in our research was fixed at 0.07. Meanwhile, $K$ symbolizes the batch size per step, with $K$ being a subset of $N$. Through the total loss, $\mathcal{L}_{\mathrm{total}}$, our model gets trained to maximize mutual information between the aligned ECG-text pairs that encompass cross-referential attributes in a batch.

\subsection{Self-supervised Contrastive Learning}
Conventional contrastive-based SSL methods \cite{simclr,byol,simsiam,tstcc,astcl} rely on strong augmentation to generate two distinct views for the input data, such as random segmentation and inversion, to the original ECG signals. This creates augmented views that serve as positive pairs $[(e_{ecg, i}, e_{ecg, i}^{{\prime}})]_{i=1}^{N}$, with the other ECG signals in the mini-batch being considered as negative examples.  The pipeline is depicted in Fig \ref{1b}. This data augmentation approach aligns with the strategy outlined in~\cite{tstcc}.

Next, we derive the representations of these augmented views, represented as $[\mathbf{\hat{e}^{\prime}}]_{i=1}^{N}$, using the ECG projector $p_{ecg}$ and ECG encoder $\mathcal{F}_{ecg}$. This is analogous to obtaining the representations $[\mathbf{\hat{e}}]_{i=1}^{N}$. Consequently, our ECG invariant learning goal is defined as:

\begin{equation}
\mathcal{L}_{SSL} = - \frac{1}{K} \sum_{j=1}^{N}{\log \frac{\mathrm{exp}(r_{i,i}^{e2e^{\prime}} / \sigma_{2})}{\sum_{j=1}^{N}{\mathrm{exp}(r_{i,j}^{e2e^{\prime}} / \sigma_{2})}}}
\label{ssl eq}
\end{equation}

\begin{align}
    \mathbf{\hat{e}}_{ecg, i} = \mathcal{F}_{ecg}(e_{ecg, i}), \mathbf{\hat{e}^{\prime}}_{ecg, i} = \mathcal{F}_{ecg}(e_{ecg, i}^{{\prime}})
\end{align}

\begin{equation}
    r_{i,i}^{e2e^{\prime}} = \mathbf{\hat{e}}_{ecg, i}^{\top} \mathbf{\hat{e}^{\prime}}_{ecg, i}
\end{equation}
In Eq \ref{ssl eq}, the temperature hyper-parameter $\sigma_{2}$ retains its value of 0.07 when considering the overall loss objective $\mathcal{L}_{SSL}$.

\section{Experiments and Analysis}

\subsection{Datasets}
\noindent \textbf{PTB-XL} The ECG dataset under examination is substantial, encompassing 21,837 ECG signals that were accumulated from 18,885 patients during the period of October 1989 to June 1996. The collected data consists of 12-lead ECG, each sampled at a rate of 500 Hz with a duration of 10 seconds, where each ECG signal is paired with the corresponding ECG reports. The reports are generated by the standard protocol and only describe the ECG without final diagnosis. The original ECG reports were written in 70.89\% German, 27.9\% English, and 1.21\% Swedish, and were converted into structured SCP-ECG statements. 
For downstream tasks, we follow the official split from \cite{astcl} to build the train/val/test split only with single category. Furthermore, each record in downstream task setting is classified under one of five primary diagnostic categories: Normal (NORM), Myocardial Infarction (MI), ST/T Change (STTC), Conduction Disturbance (CD), and Hypertrophy (HYP).

\noindent \textbf{CPSC2018} This dataset, which is publicly accessible, comprises 6,877 standard 12-lead ECG records, each sampled at a rate of 500 Hz, and the duration of these records ranges from 6 to 60 seconds. The dataset is annotated with nine distinct labels, which include Atrial fibrillation (AF), First-degree atrioventricular block (I-AVB), Left bundle branch block (LBBB), Right bundle branch block (RBBB), Premature atrial contraction (PAC), Premature ventricular contraction (PVC), ST-segment depression (STD), ST-segment elevation (STE), and normal (Normal).

For both datasets, we adhere to the official split as outlined in~\cite{astcl} and only select samples that belong to a single category.

\subsection{Implementation}
The ECG encoder we utilize is ResNet18-1D. This is adapted from its two-dimensional version, ResNet18-2D \cite{resnet}, by transitioning to 1D convolutional layers. For text encoding, we employ BioClinicalBERT \cite{alsentzer2019publicly}, pre-trained on clinical notes and bio-clinical articles. Our model integrates two linear projection heads: one for the ECG encoder and another for the text encoder. Both are characterized by an output dimension of 512 and utilize a temperature parameter $\tau$ initialized to 0.07. The ECG encoder's optimization is handled using the Adam optimizer, set with a learning rate and weight decay of $2\times e^{-3}$ and $1 \times e^{-5}$. During pre-training, we operate over 50 epochs with a batch size of 128, while all subsequent downstream tasks are processed with a batch size of 32. All experimental procedures are executed using PyTorch 2.0 on 1 NVIDIA A100-40GB GPU.

\subsection{Results on Linear Evaluation}

In the task of linear evaluation, we rigorously test the quality and robustness of the ECG representations generated by our ETP framework. To do this, we keep the pre-trained ECG encoder fixed and only update a linear classifier that is initialized randomly. This evaluation methodology is applied to two large-scale public ECG datasets with disease-level annotation, PTB-XL and CPSC2018, using Area Under the Curve (AUC) score and F1-score as the primary metrics for performance assessment. As clearly indicated in Table \ref{tab: linear res}, the ETP framework sets new performance standards, outclassing all existing baseline methods. Specifically, it achieves an AUC of 83.5 and an F1-score of 61.3 on the PTB-XL dataset. Similarly, on the CPSC2018 dataset, ETP registers an AUC of 86.1 and an F1-score of 63.4. These compelling results not only validate the effectiveness of ETP but also firmly establish it as the leading methodology for learning ECG representations.

\begin{table}[h!]
\centering
\caption{Linear evaluation results on PTB-XL and CPSC2018. Best results are in bold.}
\label{tab: linear res}
\scalebox{0.99}{
\begin{tabular}{|c|c c|c c|}
\hline \hline 
&  \multicolumn{2}{c|}{PTB-XL} & \multicolumn{2}{c|}{CPSC2018} \\
\hline 
 Method & AUC & F1 & AUC & F1 \\
\hline 
Random init & 71.5 & 52.3 & 72.1 & 59.9 \\
\hline 
CPC~\cite{cpc} & 70.3 & 54.2 & 74.6 & 53.6 \\
\hline 
SimCLR~\cite{simclr} & 67.5 & 55.5 & 73.2 & 56.8 \\
\hline 
BYOL~\cite{byol} & 76.1 & 56.8 & 77.4 & 61.3 \\
\hline 
SimSiam \cite{simsiam} & 71.4 & 56.8 & 75.5 & 62.0 \\
\hline 
TS-TCC \cite{tstcc} & 81.8 & 56.4 & 83.5 & 62.2 \\
\hline 
CLOCS \cite{clocs} & 81.7 & 55.8 & 82.0 & 61.3 \\
\hline 
ASTCL \cite{astcl} & 82.0 & 57.4 & 84.2 & 62.8 \\
\hline 
\hline 
\textbf{ETP} & $\textbf{83.5}$ & $\textbf{61.3}$ & $\textbf{86.1}$ & $\textbf{63.4}$ \\
\hline \hline
\end{tabular}
}
\end{table}

\subsection{Results on Zero-shot Classification}

To delve deeper into the capabilities of learn cross-modal representation from proposed ETP framework, we conducted zero-shot classification tasks on both PTB-XL and CPSC2018 datasets. The results are presented in Tab \ref{tab: ptb res} and \ref{tab: cpsc res}. Our zero-shot classification pipeline is inspired by the CLIP framework \cite{clip}. We employ the phrase `this ECG indicates \underline{disease name}' as a positive prompt and calculate the cosine similarity between the ECG and prompt embeddings. The prompt with the highest similarity is selected as the predicted category.\\
\textbf{PTB-XL} As shown in Table \ref{tab: ptb res}, the ETP pre-trained model consistently surpasses models with random initialization across various metrics, including AUC, ACC, and F1-score. For example, in the `NORM' category, ETP achieves an AUC of 71.8, compared to 52.7 for random initialization. It also attains an ACC of 87.4 in the `HYP' category, as opposed to 10.6 with random initialization. However, it's important to note that the AUC score for ETP is lower in the `MI' category, indicating potential areas for improvement in specific disease classifications. Overall, ETP demonstrates significant enhancements, with average scores of 54.6 for AUC, 60.8 for ACC, and 33.1 for F1-score.\\
\textbf{CPSC2018} 
Table \ref{tab: cpsc res} shows similar trends. The ETP pre-trained model consistently outperforms models initialized randomly across various metrics, such as AUC, ACC, and F1-score. Specifically, in categories like `LBBB,' ETP achieves an AUC of 81.3, compared to 33.3 from a randomly initialized model. Additionally, ETP attains an ACC of 72.3 in the `PAC' category, as opposed to 9.6 from a randomly initialized model. The average scores across all categories further underscore ETP's superiority, with an average AUC of 57.1, ACC of 60.9, and F1-score of 27.1. These substantial improvements across all disease categories highlight the effectiveness of the cross-modal representation learned by ETP.

The results affirm the efficacy of ETP in learning robust cross-modal representations for ECG and paired reports. While ETP shows promising results in most categories, there are specific areas, such as the `MI' category in the PTB-XL dataset, where further refinement could be beneficial. Overall, the ETP framework demonstrates a compelling advantage over random initialization in zero-shot classification tasks, thereby validating its potential for practical applications in cardiovascular healthcare.

\begin{table}[h!]
\centering
\caption{Zero-shot classification Results on PTB-XL. Best results are in bold.}
\label{tab: ptb res}
\scalebox{0.8}{
\begin{tabular}{|c|c|c c c|}
\hline \hline 
& & \multicolumn{3}{c|}{PTB-XL} \\
\hline 
Category & Method & AUC & ACC & F1 \\
\hline 
\multirow{2}{*}{NORM} & Random init & 52.7 & 58.8 & 72.8 \\
& \textbf{ETP} & $\textbf{71.8}$ & $\textbf{56.8}$ & $\textbf{73.4}$ \\
\hline 
\multirow{2}{*}{MI} & Random init & $\textbf{57.6}$ & $\textbf{54.4}$ & $\textbf{28.7}$ \\
& \textbf{ETP} & 46.4 & 15.5 & 26.6 \\
\hline 
\multirow{2}{*}{STTC} & Random init & 55.3 & 43.7 & 26.4 \\
& \textbf{ETP} & $\textbf{56.3}$ & $\textbf{57.8}$ & $\textbf{24.8}$ \\
\hline 
\multirow{2}{*}{CD} & Random init & 35.5 & 10.6 & 19.3 \\
& \textbf{ETP Pre-trained} & $\textbf{52.6}$ & $\textbf{87.4}$ & $\textbf{28.1}$ \\
\hline 
\multirow{2}{*}{HYP} & Random init & 25.4 & 3.5 & 6.3 \\
& \textbf{ETP} & $\textbf{45.8}$ & $\textbf{86.4}$ & $\textbf{12.2}$ \\
\hline 
\multirow{2}{*}{Average} & Random init & 45.3 & 34.2 & 30.7 \\
& \textbf{ETP} & $\textbf{54.6}$ & $\textbf{60.8}$ & $\textbf{33.1} $ \\
\hline \hline
\end{tabular}
}
\end{table}

\begin{table}[h!]
\centering
\caption{Zero-shot classification Results on CPSC2018. Best results are in bold.}
\label{tab: cpsc res}
\scalebox{0.9}{
\begin{tabular}{|c|c|c c c|}
\hline \hline 
& & \multicolumn{3}{c|}{CPSC2018} \\
\hline 
Category & Method & AUC & ACC & F1 \\
\hline 
\multirow{2}{*}{Normal} & Random init & 51.6 & 23.1 & 21.9 \\
& \textbf{ETP} & $\textbf{55.1}$ & $\textbf{60.6}$ & $\textbf{26.8}$ \\
\hline 
\multirow{2}{*}{AF} & Random init & 49.9 & 15.9 & 27.4 \\
& \textbf{ETP} & $\textbf{50.9}$ & $\textbf{52.7}$ & $\textbf{32.5}$ \\
\hline 
\multirow{2}{*}{I-AVB} & Random init & 46.3 & 47.4 & 22.7 \\
& \textbf{ETP} & $\textbf{50.8}$ & $\textbf{51.8}$ & $\textbf{23.4}$ \\
\hline 
\multirow{2}{*}{LBBB} & Random init & 33.3 & 4.9 & 6.0 \\
& \textbf{ETP} & $\textbf{81.3}$ & $\textbf{94.0}$ & $\textbf{35.1}$ \\
\hline 
\multirow{2}{*}{RBBB} & Random init & 45.3 & 24.3 & 39.3 \\
& \textbf{ETP} & $\textbf{55.3}$ & $\textbf{24.3}$ & $\textbf{39.3}$ \\
\hline 
\multirow{2}{*}{PAC} & Random init & 39.9 & 9.6 & 16.4 \\
& \textbf{ETP Pre-trained} & $\textbf{46.3}$ & $\textbf{72.3}$ & $\textbf{18.8}$ \\
\hline 
\multirow{2}{*}{PVC} & Random init & 43.8 & 10.8 & 19.4 \\
& \textbf{ETP} & $\textbf{65.9}$ & $\textbf{78.3}$ & $\textbf{27.0}$ \\
\hline 
\multirow{2}{*}{STD} & Random init & 39.5 & $\textbf{35.1}$ & 22.7 \\
& \textbf{ETP} & $\textbf{47.0}$ & 19.6 & $\textbf{24.3}$ \\
\hline 
\multirow{2}{*}{STE} & Random init & 45.0 & 27.0 & 8.1 \\
& \textbf{ETP} & $\textbf{61.2}$ & $\textbf{95.0}$ & $\textbf{16.2}$ \\
\hline 
\multirow{2}{*}{Average} & Random init & 43.8 & 24.2 & 18.2 \\
& \textbf{ETP} & $\textbf{57.1}$ & $\textbf{60.9}$ & $\textbf{27.1}$ \\
\hline \hline
\end{tabular}
}
\end{table}

\section{Conclusion}
In this work, we propose ETP, the novel framework to learn cross-modal representation from unannotated ECG and associated report. We also first build the comprehensive benchmark on linear evaluation and zero-shot classification with ECG cross-modal learning and SSL. ETP surpass all SSL methods on linear evaluation task and endow the zero-shot ability to ECG community via the proposed framework and evaluated on two large-scale public datasets, PTB-XL and CPSC2018. Overall, this work establishes the first comprehensive benchmark for ECG zero-shot classification and cross-modal learning, demonstrating the capability and potential of jointly learning ECG signals and paired  reports.



\clearpage
\small
\bibliographystyle{IEEE.bst}
\bibliography{refs.bib}
\end{document}